\RequirePackage[mathlines]{lineno} 
\documentclass[a4paper,11pt]{article}
\pdfoutput=1 

\usepackage{jheppub} 

\usepackage[T1]{fontenc} 
\usepackage{comment}
\usepackage{hyperref}
\hypersetup{hidelinks,
	colorlinks=true,
	allcolors=blue,
	pdfstartview=Fit,
	breaklinks=true}
\usepackage{threeparttable}
\usepackage{multirow}
\usepackage{subfigure}
\usepackage{float}
\let\oldequation\equation
\let\oldendequation\endequation
\renewenvironment{equation}
  {\linenomathNonumbers\oldequation}
  {\oldendequation\endlinenomath}

\def \ee   {e^+e^-}

\def\effnuminv{(17.14\pm 0.04)\%}
\def\effnumtag{(16.02 \pm 0.06)\%}

\def\bfinv{8.4\times 10^{-4}}

\def\eemc{\textit{E}_{\rm{EMC}}}
\def\ksany{K_S^0\to\rm{anything}}

\def \gevcc{~\mbox{GeV/$c^2$}}

\def\eemc{ E_{\rm{EMC}}}
\def\BR{\mathcal{B}}
\def\jpsi{J/\psi}
\def\Ks{K_S^0}
\def\KL{K_L^0}
\def \tagmode{J/\psi \rightarrow\phi K_{S}^0 K_{S}^0}
\def \Tagmode{J/\psi \rightarrow\phi K_{S}^0(\text{tag}) K_{S}^0}
\def \phiksks {\phi K_S^0 K_S^0}
\def\ee{e^+e^-}

\def\efftag{\varepsilon_{\rm{non}-\pi^+\pi^-}}

\def\effinv{\varepsilon_{\text {signal}}}

\def\rt{\rightarrow}

\def\kspipi{K_S^0\rightarrow \pi^+\pi^-}
\def\kkksks{J/\psi\to K^+ K^- K_S^0({\rm{tag}})K_S^0}

\def\phikk{\phi \rightarrow K^+K^-}
\def\Ntag{N_{\rm{non}-\pi^+\pi^-}}
\def\Ninv{N_{\rm{signal}}}
\def\Kstag{K_{S}^{0}(\rm{tag})}
\def \num {\mbox{(1.0087 $\pm$ 0.0044) $\times 10^{10}$}}

\def\ksinv {K_S^0\rightarrow \rm{invisible}}
\def\ksany {K_S^0\rightarrow \rm{anything}}

\def\pippim{\pi^+\pi^-}

\begin{document}

\title{\boldmath Search for $K^0_S$ invisible decays}

\collaborationImg{\includegraphics[height=30mm,angle=90]{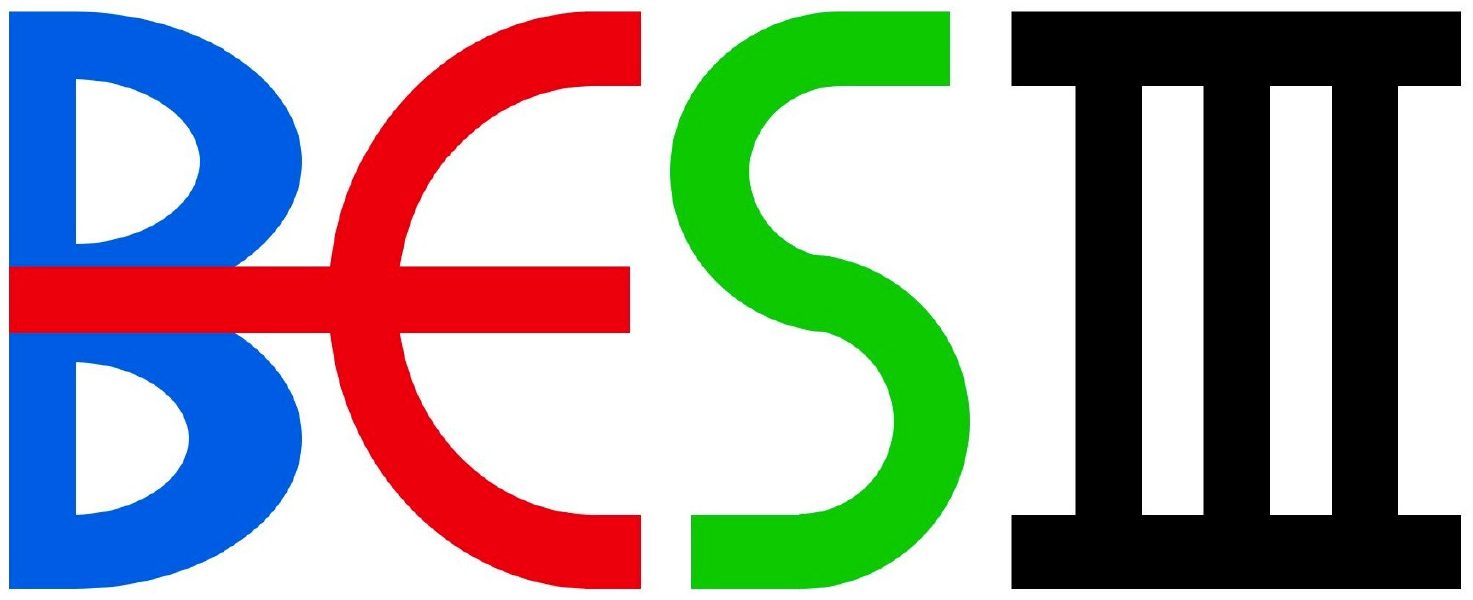}}
\collaboration{The BESIII collaboration}
\emailAdd{besiii-publications@ihep.ac.cn}

\abstract{
Based on $(1.0087\pm0.0044)\times10^{10}$ $J/\psi$ events collected with the BESIII detector at the BEPCII $e^+e^-$ storage ring, we search for $\Ks$ invisible decays via the $\tagmode$ process. No significant signal is observed, and the upper limit of the branching fraction of these invisible decays is set at  8.4 $\times$ $10^{-4}$ at the  90\% confidence level. This is the first experimental search for $K^0_S$ invisible decays.
}



\maketitle
\flushbottom

\section{Introduction}
\label{sec:introduction}
\hspace{1.5em} 
The Standard Model (SM) of particle physics has been a cornerstone in understanding the subatomic world over the past several decades, providing a comprehensive framework for explaining many  observed phenomena.   Despite its extensive successes, the SM does not address  certain issues, most notably dark matter~\cite{NP}.
 Accumulating indirect evidence
from astronomical and cosmological observations strongly suggests  the existence of dark matter~\cite{reviewDM, rotationcurve}, which is invisible in the entire electromagnetic spectrum, and its existence is inferred via gravitational effects only.  Studies of invisible decays, where particles decay into final states  that do not produce detectable signals, are therefore important for the development of SM extensions~\cite{invdecay, invdecay2}.

Stringent limits on the invisible decays of the $\Upsilon$~\cite{Upsiloninv}, $J/\psi$~\cite{Jpsiinv},  $B^0$~\cite{B0inv}, $B_s$~\cite{Bsinv},  $\eta(\eta')$~\cite{etainv,etainv_bes2}, $\pi^0$~\cite{pi0inv}, $D^0$~\cite{D0inv}, $\omega$~\cite{omgphi}, $\phi$~\cite{omgphi} mesons and the $\Lambda$~\cite{lmdinv} baryon have already been set by several
experiments. However, no experimental study of fully-invisible decays of  kaons has been performed yet. 
 Within the SM, the branching fraction (BF) of $K^0_S\to \nu\bar \nu$ decay is predicted to be extremely small. This process is kinematically forbidden under the assumption of  massless neutrinos due to  angular momentum conservation, and remains highly suppressed in the case of massive neutrinos   due to the unfavorable helicity configuration, with a BF smaller than $10^{-16}$~\cite{ULks}. Consequently, the search of the $\Ks$ invisible decay offers a sensitive test of the SM~\cite{invdecay2}.

By summing  all the known $\Ks$  decay modes, an indirect estimation of the BF allowing $\Ks$ to decay invisibly is established at the order of $10^{-4}$~\cite{ULks}.
 Additionally, theories  like the mirror-matter model~\cite{Mirrormodel,Mirro2}, which assumes   the existence of  a mirror world parallel to our own, suggest that the  $K^0_S$ invisible decay 
 could be interpreted as an oscillation between normal and mirror particles, and predict the BF of $\ksinv$ to be at the order of   $10^{-6}$. 
As there has been no experimental exploration of 
$\ksinv$ reported,  the indirect experimental upper limit (UL) and the model prediction both remain unverified.

Study of the $K^0_S$ invisible decay 
is also essential for testing CPT invariance~\cite{ULks}.  Using the the neutral kaon system  for such tests offers advantages over the  $B^0$ or $D^0$ meson systems; specifically, one benefits from the small total decay widths and the limited number of significant (hadronic) decay modes~\cite{CPT}.  The Bell-Steinberger relation (BSR)~\cite{BSR}, derived from the requirement of unitarity, connects potential CPT-invariance violation to the amplitudes of all decay channels of neutral kaons.  
Although the BSR provides the most sensitive test of CPT symmetry, previous BSR tests with neutral kaons have been conducted assuming that there is no contribution from invisible decay modes.

 In this paper, we report the first experimental search for $\Ks$ invisible decays via the $\tagmode$ decay, by analyzing  $\num$ $\jpsi$ events collected with the BESIII detector at the BEPCII $e^+e^-$ storage ring~\cite{bes3:njpsi2022}. The usage of   $\tagmode$  provides a unique advantage for probing $\Ks$ invisible decays. Most $\jpsi$ decay modes with $\Ks$ in the final states   suffer from high contamination from $\KL$ background, which can mimic the  signal. In contrast, in for $\tagmode$ decay, with a BF of ($5.9\pm 1.5$)$\times 10^{-4}$~\cite{pdg:2024}, one of the dominant $\KL$ backgrounds, $\jpsi\to \phi \Ks\KL$ is forbidden by C-parity conservation. This    enables us to probe the $K^0_S$ invisible decay 
 signal from a relatively clean $\Ks$ sample.

\section{BESIII detector and Monte Carlo simulation}
\label{sec:detector}
\hspace{1.5em}
The BESIII detector~\cite{Ablikim:2009aa} records symmetric $e^+e^-$ collisions 
provided by the BEPCII storage ring~\cite{Yu:IPAC2016-TUYA01} in the center-of-mass energy range from 1.84 to 4.95~GeV, with a peak luminosity of $1.1\times10^{33}\;\text{cm}^{-2}\text{s}^{-1}$ achieved at $\sqrt{s} = 3.773\;\text{GeV}$.
BESIII has collected large data samples in this energy region~\cite{Ablikim:2019hff}. The cylindrical core of the BESIII detector covers 93\% of the full solid angle and consists of a helium-based
 multilayer drift chamber~(MDC), a plastic scintillator time-of-flight
system~(TOF), and a CsI(Tl) electromagnetic calorimeter~(EMC),
which are 
all enclosed in a superconducting solenoidal magnet providing a 1.0~T magnetic field.
The magnetic field was 0.9~T in 2012, which affects 11\% of the total $J/\psi$ data.
The solenoid is supported by an
octagonal flux-return yoke with resistive plate counter muon
identification modules (MUC) interleaved with steel. 

The charged-particle momentum resolution at $1~{\rm GeV}/c$ is
$0.5\%$, and the specific ionization energy loss (d$E$/d$x$) resolution is $6\%$ for electrons
from Bhabha scattering. The EMC measures photon energies with a
resolution of $2.5\%$ ($5\%$) at $1$~GeV in the barrel (end-cap)
region. The time resolution in the TOF barrel region is 68~ps, while
that in the end-cap region is 110~ps. 
The end-cap TOF system was upgraded in 2015 using multi-gap resistive plate chamber
technology, providing a time resolution of 60~ps, which benefits 87\% of the data used in this analysis~\cite{etof1, etof2}.

Simulated data samples produced with the {\sc geant4}-based~\cite{geant4} Monte Carlo (MC) package, which includes the geometric and material description of the BESIII detector~\cite{detvis,geo1,geo2} and the detector response, are used to determine detection efficiencies and to estimate backgrounds. The simulation models the beam energy spread and initial state radiation in the $e^+e^-$ annihilations with the generator {\sc kkmc}~\cite{ref:kkmc1, ref:kkmc2}. 
The inclusive MC sample includes the production of the $J/\psi$ resonance incorporated in {\sc kkmc}.
All particle decays are modeled with {\sc evtgen}~\cite{ref:evtgen1, ref:evtgen2} using the BFs either taken from the Particle Data Group~(PDG)~\cite{pdg:2024}, when available, or otherwise estimated with {\sc lundcharm}~\cite{ref:lundcharm1, ref:lundcharm2}. Final state radiation from charged final state particles is incorporated using the {\sc photos} package~\cite{photos}.
The signal MC sample for $\tagmode$ is generated using a phase space model. To enhance the accuracy of the signal model, a multidimensional re-weighting method as described in Ref.~\cite{reweight} is employed. Detailed information about this re-weighting method is provided in Sec.~\ref{sec:evt_select}.

\section{Analysis method}
\label{sec:analysis}
\hspace{1.5em} 
In this analysis, the  $\Ks$ sample is selected using the  $\tagmode$ process. To study the $\ksinv$ without relying on the BF of $\tagmode$, which suffers from significant uncertainties, a novel method is employed. In this method, we define a non-$\pippim$ sample first,  containing the events that satisfy $\tagmode$, $\phikk$, $\kspipi$   with the $\Ks$ in the recoiling system  decaying to processes other than $\pippim$. The $\Ks$ decaying to $\pippim$ is denoted as $\Kstag$ hereafter.   In such cases, each selected event inherently qualifies as a candidate for the $\Ks$ invisible decay, since  the non-$\pippim$ event only contains four charged particles. Subsequently, we can probe the $\ksinv$ decay using the identical dataset, where  the $\ksinv$ candidate is searched for in the system recoiling against a reconstructed $\phi\Ks$ candidate.

The yields for the selected non-$\pippim$ sample and the $\ksinv$ signal events  are denoted as $\Ntag$ and $\Ninv$, which are given by:
\begin{align}
    \Ntag &= 2\times  N_{\tagmode}\times \BR(\phikk)\times \BR(\kspipi) \nonumber \\
    & \times (1-\BR(\kspipi)) \times \efftag, \label{eq:tag} 
\end{align}
and
\begin{align}
   \Ninv &= 2\times N_{\tagmode}\times \mathcal{B}(\phikk) \times \BR(\kspipi) \nonumber \\
  & \times \mathcal{B}(K_S^0 \rightarrow  \text{invisible}) \times  \effinv, \label{eq:vis}
\end{align}
respectively. Here, 
 $N_{\tagmode}$ represents the product  of the total number of $\jpsi$ events and the BF of $\tagmode$, while $\BR(\phikk)$ and $\BR(\kspipi)$ are the   BFs of  $\phikk$ and $\kspipi$  quoted from the PDG~\cite{pdg:2024}, respectively.  The term $1-\BR(\kspipi)$ stands for the probability that $\Ks$ decays to processes other than $\pippim$, which corresponds to our definition of the non-$\pippim$ sample.  The efficiencies  of  selecting   the non-$\pippim$ sample and  the signal event are denoted by  $\efftag$ and $\effinv$, respectively.
The  BF of the $\ksinv$  decay is determined as:
\begin{equation}
\mathcal{B}(K_S^0 \rightarrow \text { invisible})= \frac{\Ninv}{ \Ntag \, (\effinv/\efftag)} \, (1-\BR(\kspipi)).
\label{eq:relative_ratio}
\end{equation}

In this approach, the systematic
uncertainties arising from the total number of $\jpsi$ events, the BFs of $\tagmode$ and $\phikk$ cancel, and that from the reconstruction efficiency mostly cancels. To avoid a possible bias,  a semi-blind analysis is conducted using 10\% of the full data sample to validate the analysis strategy. The results presented herein are derived from the full data sample,  with the analysis method predetermined and  fixed from the 10\% sample.


\section{Event selection and data analysis}
\label{sec:evt_select}
\hspace{1.5em} 
To select the candidates for the non-$\pippim$ sample, where $\tagmode$, $\phikk$, and only one $\Ks$ decays to $\pippim$, we reconstruct the events 
  with exactly four charged tracks, ensuring that no additional charged track is  present. 
Charged tracks detected   in the MDC are required to be within a polar angle ($\theta$) range of $|\cos\theta|<0.93$, where $\theta$ is defined with respect to the $z$ axis, which is the symmetry axis of the MDC. For  
charged tracks originating from $\phi$    decays, the distance of the closest approach to the  interaction point (IP), |$V_z$|, 
must be less than 10\,cm along the $z$ axis and less than 1\,cm in the  plane perpendicular to the
z axis.  Particle identification~(PID) for charged tracks is implemented by combining measurements of the d$E$/d$x$  in the MDC and the flight time in the TOF to form likelihoods $\mathcal{L}(h), h=K,\pi$, for each hadron $h$ hypothesis.
Charged tracks are 
 identified as kaons by requiring
  $\mathcal{L}(K)>\mathcal{L}(\pi)$. The $\phi$ meson is reconstructed through the decay $\phikk$, and its invariant mass, $M(K^+ K^-)$, is required to be in the range of $[1.00 , 1.04]\gevcc$.
 
  The $\Kstag$ candidates are reconstructed using two oppositely charged tracks, which are each required to satisfy $|V_z|$ < 20\,cm.  Tracks are then identified as pions  by requiring   $\mathcal{L}(\pi)>\mathcal{L}(K)$. 
A vertex fit constraints the $\pi^+\pi^-$ pairs are constrained to originate from a common vertex.
A further fit then constrains the momentum of the $\Kstag$ candidate 
to point from the IP to the decay vertex.  
The decay length of the $\Kstag$ candidate is required
to be greater than twice the vertex resolution. The signal
  region for the  $\pippim$ invariant mass is $0.486 \gevcc <M(\pippim)<0.510\gevcc$.

To further suppress the background  from $\jpsi\to \gamma\phi\phi$, $\phikk$, $\phi\to \Ks\KL$, the recoil mass of the  selected $\phi$ candidate is required to be greater than 1.08$\gevcc$.
In addition, we require  the cosine of the 
 polar angle of the  $\phi\Kstag$ system to be within the interval [$-0.80, 0.80$]. This condition ensures most of the decay products of the recoiling $\Ks$ fall within the acceptance region of the barrel EMC. Furthermore, the recoil mass of the selected $\phi\Kstag$ candidate must  be within 40 MeV/$c^2$ of the known $\Ks$ mass~\cite{pdg:2024}.

\vspace{-0.0cm}
\begin{figure}[htbp] \centering
	\setlength{\abovecaptionskip}{-1pt}
	\setlength{\belowcaptionskip}{10pt}
	\includegraphics[width=10.0cm]{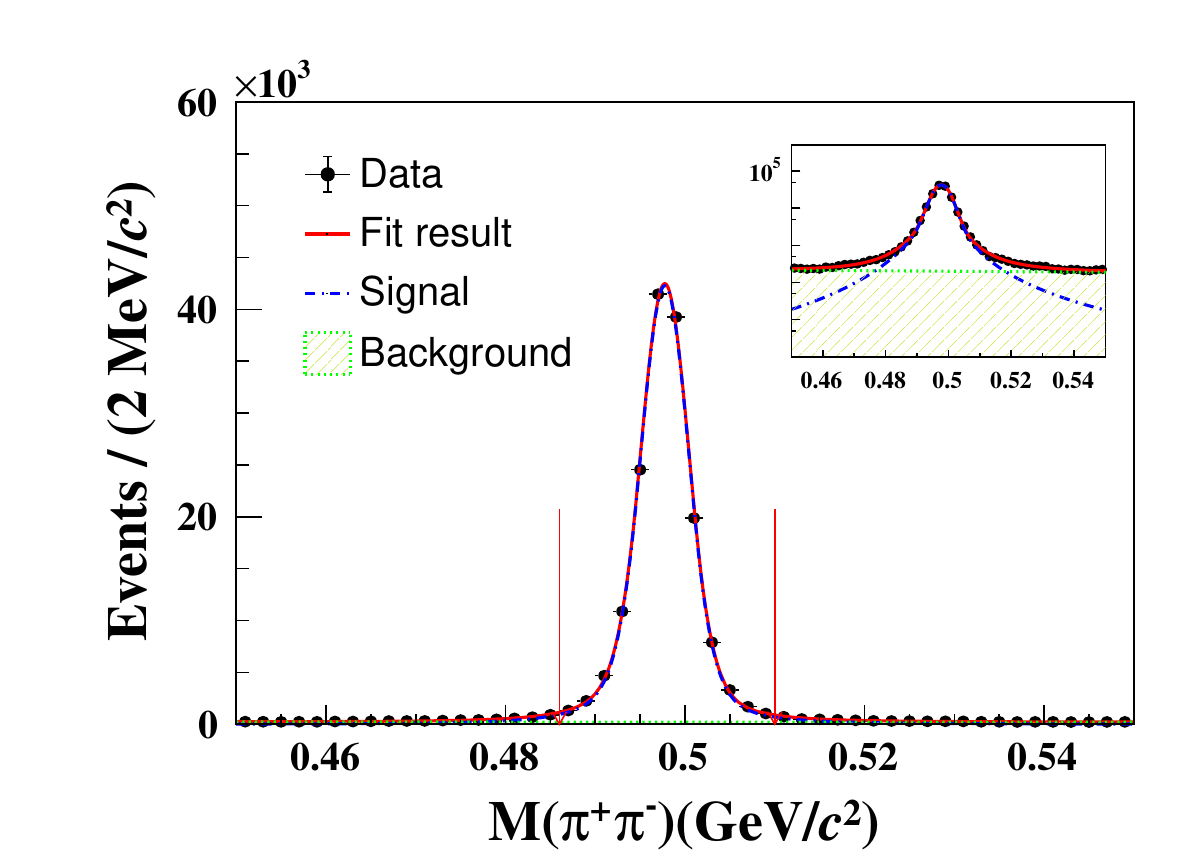}
	\caption{ Fit to the distribution of $M(\pi^+\pi^-)$. The black dots with error bars represent data and the red solid line shows the total fit. The blue curve and green dashed curve are the fitted signal and background shapes, respectively. The red arrows denote the $K^0_S$ signal region. The inset of the figure displays the fit result with a logarithmic vertical scale.}
	\label{fig:fit}
\end{figure}

After applying all  the above selection requirements,  the analysis of the $\jpsi$  inclusive MC sample 
indicates that the remaining backgrounds affecting $\Ntag$  can be categorized into two types: the four-pion and non-$\phi$ background. The four-pion background primarily originates from $\tagmode$, with both  $\Ks$ mesons decaying to $\pippim$. The expected yield of the four-pion background in data, as estimated from  MC simulation and normalized to the full data sample, is 1022 $\pm$ 260.
The non-$\phi$  background is  from $\jpsi\to K^+ K^-  K_S^0({\rm{tag}}) K_L^0$ and $\jpsi\to K^+ K^- K_S^0({\rm{tag}})\Ks$. 
 While the contribution from the latter decay can be estimated using MC simulation, the contribution from  $\jpsi\to K^+ K^-  K_S^0({\rm{tag}}) K_L^0$   remains uncertain because its BF has not been measured.  Therefore, at this stage, we are unable to directly estimate the contribution from the non-$\phi$ background.
Details about the estimation of this contribution will be discussed  in Sec.~\ref{sec:bkgana}.

\vspace{-0.0cm}

To extract the yield of the non-$\pippim$ sample, a binned maximum likelihood fit is performed on the distribution of  $M(\pippim)$, as depicted in Fig.~\ref{fig:fit}. 
In the fit, the signal is modeled using a double Gaussian function, while the non-peaking background is described by a second-order Chebyshev function.  The  yield   $N$ is determined to be
$(1.535\pm 0.004)\times 10^5
$ by integrating the signal function over the $K^0_S$ signal region. It is noted that $N$  represents  a preliminary   yield. Given that the four-pion and non-$\phi$ background peak in the signal region of the $M(\pippim)$ distribution,  it is necessary to subtract these contributions from $N$ to obtain the final  yield of the non-$\pippim$ sample, $\Ntag$.  

The efficiency of selecting the non-$\pippim$ sample  is determined from a MC sample of $\tagmode$,  with $\phikk$ and $\Ks\to$ inclusive.  For this MC sample, the  efficiency is obtained  by counting the  number of events that survive the selection criteria,    and   truth information is employed to  identify the events where only one $\Ks$ decays to $\pi^+\pi^-$.  To  improve the accuracy of the MC model 
for  $\tagmode$, the MC events are corrected based on  a
multidimensional re-weighting method as described in  Ref.~\cite{reweight}.   A clean control sample of $\tagmode$, with $\phikk$ and both $\Ks$ mesons decaying to $\pippim$, is selected using the selection criteria similar to that of Ref.~\cite{bam547}. Correction factors are derived from this control sample as a function of the invariant masses  of $\Ks\Ks$ and $\phi\Ks$, as well as the  cosines of the polar angles for the $\phi$ and $\Ks$, denoted as $\cos\phi$ and $\cos\Ks$. These correction factors are subsequently applied to the MC samples to correct the MC-simulated shapes, thus enabling accurate determination of the detection efficiencies for both non-$\pippim$ and signal events.  The efficiency is determined to be $\efftag =\effnumtag$, where the uncertainty comes from  MC statistics. Note
the efficiencies do not include the BFs of $\phi$ and $\Ks$ subsequent decays.

We search for the  $\ksinv$ signal using  the  same  selection criteria  as those used for selecting the non-$\pippim$ sample.  The detection efficiency $\effinv$ is determined to be $\effnuminv$ based on the signal MC sample of $\Tagmode$, $\ksinv$.  
As the $K^0_S$ invisible decay does not deposit any energy in the EMC, 
the sum of energies of all EMC showers not associated with any charged tracks, 
$\eemc$,  can be used to distinguish the signal from background. 
 For the selected  showers, we require that they are  separated by more than $10^{\circ}$
 from other charged tracks, and the  difference between the EMC time and the event start time is required to be within 700 ns.  These requirements remove charged-particle showers and help suppress electronic noise and showers unrelated to the event.

\section{Background analysis}
\label{sec:bkgana}
\hspace{1.5em} 
The dominant backgrounds  affecting the signal side yield, $\Ninv$, arise from the following three sources:
\begin{itemize}
    \item  $\ksany$ background, which comes from $\Tagmode$, with $\Ks$ decaying to visible particles,  such as $\Ks\to \pi^0\pi^0$ and $\kspipi$.
 Notably, the background  from $\Ks$ decaying to charged particles, like $\kspipi$,  is strongly suppressed by the selection requirements for four-charged tracks and are thus    negligible   compared to   $\Ks\to \pi^0\pi^0$.   Consequently,  the energy deposited in the EMC ($E_{\rm EMC}$) of $\ksany$ background is primarily studied using the control sample of  $\Tagmode$, $\Ks\to \pi^0\pi^0$. Good consistency in the $E_{\rm EMC}$ distributions between data and MC simulation allows  us to model the $\ksany$ background using the  MC-simulated shape based on the  $\Ks\to \pi^0\pi^0$ control sample.
    \item Non-$\phi$ background, which  originates from $\kkksks$  and $J/\psi\to K^+ K^-  K_{S}^{0}({\rm{tag}}) K_L^0$ . 
The $E_{\rm EMC}$ of the  non-$\phi$ background is characterized by that of the sideband region of  $M(K^+ K^-)$ in data, defined as $1.10 \gevcc <M(K^+ K^-)< 1.14 \gevcc$. 
The shape remains stable when using alternative sideband region, and  the   impact of the sideband choice will be
considered as a source of systematic uncertainty.
    \item Other  backgrounds, which arise  from     $\jpsi$ decays, such as    $\jpsi\to\pippim\eta K^+K^-$, and from the continuum process, i.e, $\ee\to\phiksks$. The former is studied based on the  inclusive MC sample, while the latter is assessed with continuum data collected at  $\sqrt{s}$ $=$ 3.08 GeV. 
  
\end{itemize}
\vspace{-0.0cm}
\begin{figure}[htbp] \centering
	\setlength{\abovecaptionskip}{-1pt}
	\setlength{\belowcaptionskip}{10pt}
	\includegraphics[width=10.0cm]{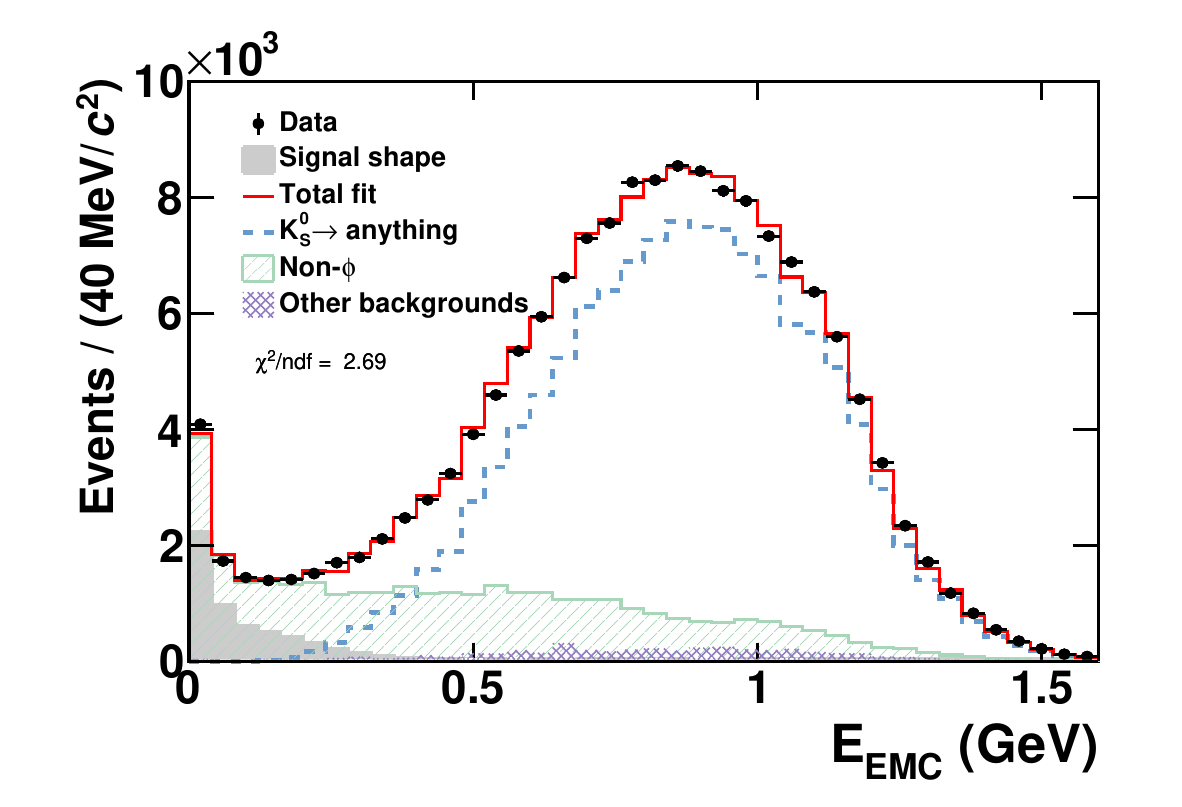}
	\caption{The distributions of $\eemc$ for the accepted candidates in data  and the inclusive MC sample. The black dots with error bars are data,  the green shaded histogram   shows the  non-$\phi$ background, and the purple shaded histogram  shows
the  other backgrounds in inclusive MC sample. The blue line shows the $\ksany$ background and the   red solid line shows the total fit. The
gray shaded  histogram shows the signal shape, normalized to  a BF of 8.0$\times 10^{-3}$ (10x the obtained limit). }
	\label{fig:fit_eemc}
\end{figure}
\vspace{-0.0cm}

The  $\eemc$  distribution, as shown in Fig.~\ref{fig:fit_eemc}, demonstrates that the total $\eemc$ distribution from the background model agrees well with the data.
A binned maximum likelihood fit is performed to determine the signal yield. In the fit, the signal is described by the  MC-simulated shape, which is corrected based on the $\kspipi$ control  sample as detailed in Sec.~\ref{sec:evt_select}. The  yield  of the non-$\phi$ background is  a free parameter in the fit,  while the total contribution from both non-$\phi$ and $\ksany$ backgrounds is fixed to the preliminary  yield, $N$.    The continuum background is characterized using the shape derived from  the continuum data  at $\sqrt{s}$ = 3.08 GeV, with a yield normalized to the
$\jpsi$ data sample, after taking into account different integrated luminosities and
center-of-mass energies~\cite{bes3:njpsi2022}. The other  backgrounds are modeled using  the shape derived   from the inclusive MC sample,  with  a yield normalized to the total number of $\jpsi$ events. 
The fit gives the signal yield $\Ninv$ to be 56 $\pm$ 201, which is consistent with zero. Additionally, the fit quantifies the  contribution from the non-$\phi$ processes, which allows determination of the final  yield for non-$\pippim$ sample by subtracting  the identified non-$\phi$  and four-pion background components from the preliminary   yield  $N$. Specifically, the contributions subtracted for the non-$\phi$ and four-pion backgrounds are $(3.457 \pm 0.049) \times 10^4$ and 1022 $\pm$ 260, respectively. The resulting  yield  is calculated to be $\Ntag$ = 
$(1.179 \pm 0.007) \times 10^5$.

Since no significant signal  is observed, an UL on the BF of $\ksinv$ is estimated after taking into account the systematic uncertainties described in the following section.

\section{Systematic uncertainties}
\label{sec:systematic}
\hspace{1.5em}
The strategy of this analysis effectively cancels many potential systematic uncertainties. Specifically, the uncertainties related to the total number of $\jpsi$ events,  $\BR(\tagmode)$ and $\BR(\phikk)$ completely cancel, and those from the selection criteria and the MC model of $\tagmode$ are greatly reduced by the ratios in Eq.~\ref{eq:relative_ratio}. 
The remaining systematic uncertainties on $\BR(\ksinv)$,  as summarized in Table~\ref{tab:syst_err}, are described below.
 

\begin{itemize}

 \item \emph{$\Ntag$.} 
To estimate the systematic uncertainty in the 
determination of $\Ntag$, we replace the signal shape of a double Gaussian with a MC-simulated shape convolved  with a Gaussian function, and vary the nominal bin size of 2 MeV/$c^2$ to either 1 MeV/$c^2$ or 3 MeV/$c^2$.
  The   maximum change in the   signal  yield,  0.7\%, is    assigned as the   systematic uncertainty.

\item \emph{BF of $\kspipi$}. The uncertainty of $\BR(\kspipi)$ is 0.1\%~\cite{pdg:2024}.

\item \emph{Signal shape.} The systematic  uncertainty due to the signal shape in the fit to $\eemc$ is evaluated  by replacing the nominal signal shape with two alternative models.
The first  model uses the   MC-simulated  shapes that are re-weighted  following the same procedure as in the nominal analysis.  The major difference, however, lies in the derivation of the correction factors, which are now functions of the momentum of $\Ks $ and $\phi$ (denoted as $p_{\Ks}$ and $p_{\phi}$), and the  cosine of the corresponding polar angles, $\cos(\phi)$ and $\cos(\Ks)$.
The second model employs the data-driven generator BODY3~\cite{bam547}, which was developed to model contributions from different intermediate states observed in data for a three-body final state. The Dalitz plot from data, corrected for backgrounds and efficiencies, is taken as input for the BODY3 generator.

    \item \emph{$\Ks\to$ anything background shape.}  To account for the uncertainty arising from the background shape of $K_S^0\rightarrow \rm{anything}$ in the fit to $\eemc$,  we employ the same alternative models as used for estimating the uncertainty related to the  signal shape. 

     \item \emph{Non-$\phi$ background shape.} 
In order to estimate the systematic uncertainty associated with the background shape of the non-$\phi$ process, alternative sideband regions, specifically [1.12, 1.16]$\gevcc$ and [1.08, 1.12]$\gevcc$, are taken into consideration.
\end{itemize}

When estimating the BF  of $\ksinv$, the correlations among different systematic uncertainties are taken into account and    varied   simultaneously  in the likelihood fit.

\begin{table*}[tpb]
\setlength{\abovecaptionskip}{0.0cm}
\setlength{\belowcaptionskip}{-1.6cm}
\caption{The systematic uncertainties in setting  the UL of  $\BR(\ksinv)$.
     The nominal analysis criteria are also included as a choice for the last three rows, but are omitted here to save space.}
  \begin{center}
  \footnotesize
  \newcommand{\tabincell}[2]{\begin{tabular}{@{}#1@{}}#2\end{tabular}}
  \begin{threeparttable}
  \begin{tabular}{l|c c c c c c c }
      \hline\hline
                         Source & Choice or uncertainty  \\
			\hline
   $\Ntag$ & 0.7\% \\
   $\BR(\kspipi)$ &   0.1\% \\
	   Signal shape    &  $\mathcal{W}(p_{\Ks}, p_{\phi}, \cos\phi, \cos K_{S}^{0})$, BODY3 MC  \\
$K_S^0\rt\rm{anything}$ background shape  &  $\mathcal{W}(p_{\Ks}, p_{\phi}, \cos\phi, \cos K_{S}^{0})$, BODY3 MC  \\ 
Non-$\phi$ background shape   &  [1.12, 1.16], [1.08, 1.12]$\gevcc$  \\
      \hline\hline
  \end{tabular}
  \label{tab:syst_err}
  \end{threeparttable}
  \end{center}
\end{table*}

\section{Result}
\label{sec:result}
\hspace{1.5em}
We employ a modified frequentist approach as  described in Refs.~\cite{Dtogenu, lmdinv, Dtopinunu}, to set the  UL of  $\BR(\ksinv)$ in Eq.~\ref{eq:relative_ratio} incorporating all the systematic and statistical uncertainties.
Thousands of toy samples are generated according to the $\eemc$ distribution observed in data. In each toy sample, the number of events is sampled from a Poisson distribution with a mean value corresponding to the data. 

For each toy sample, 
the same fit procedure used for data is performed,  where different systematic uncertainties are randomly varied.  The shapes of  signal and backgrounds, as listed in Table~\ref{tab:syst_err}, are randomly selected during the fit process.
 The total contributions from  non-$\phi$ and $\ksany$ are fixed to  the  values constrained by a Gaussian distribution, with the central value of $N$,  and the uncertainty  corresponding to the standard deviation.
 The uncertainties related to the continuum process and the other background are found to be negligible. 
 To calculate $\BR(\ksinv)$ in Eq.~\ref{eq:relative_ratio}, the  $\efftag$ and $\effinv$ are Gaussian-constrained by their respective statistical uncertainties.
 $\Ntag$ is also  Gaussian-constrained, with widths obtained by the quadrature of the statistical and systematic uncertainties, as detailed in Table~\ref{tab:syst_err}. 

\begin{figure}[htbp] \centering
	\setlength{\abovecaptionskip}{-1pt}
	\setlength{\belowcaptionskip}{10pt}
	\includegraphics[width=10.0cm]{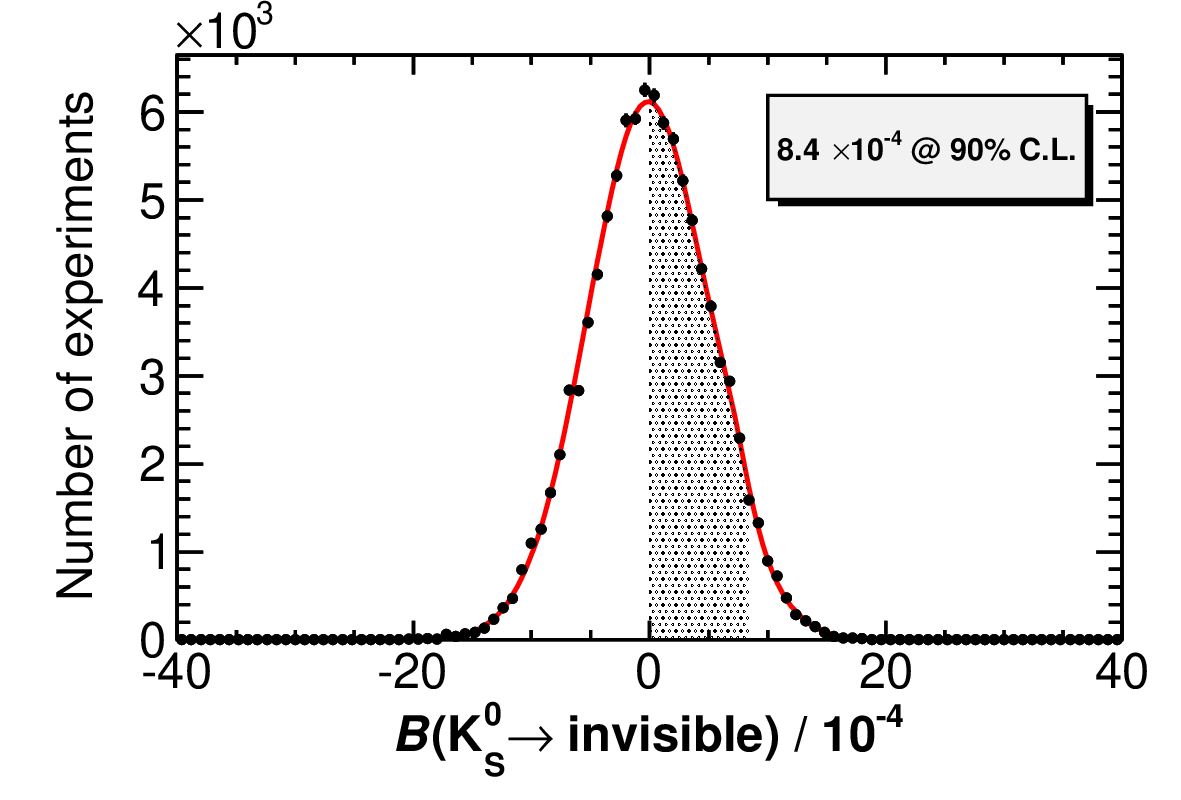}
	\caption{The distribution of  $\BR(\ksinv)$ determined from toy MC samples. The shaded area corresponds to the 90\% coverage in the physical region. }
	\label{fig:ul}
\end{figure}

The resulting distribution of the calculated  $\BR(\ksinv)$ across these toy samples is shown in Fig.~\ref{fig:ul}, which follows a Gaussian distribution as expected.  By integrating the Gaussian distribution in the physical region  greater than zero, the UL of $\BR(\ksinv)$ is determined to be $\bfinv$ at the 90\% confidence level.

\vspace{-0.0cm}
\vspace{-0.0cm}

\section{Summary}
\label{sec:summary}
\hspace{1.5em}
Based on $(1.0087$ $\pm$ 0.0044)$\times10^{10}$ $\jpsi$ events collected with the BESIII detector, we search for $\Ks$ invisible decays for the first time. 
No significant signal is observed.
The UL on the decay BF is set to be $\bfinv$ at the 90\% confidence level.  This work provides the first direct measurement of the BF   of  $K_S^0 \to \rm{invisible}$, with results that are compatible with the indirect estimation.  This search also provides a direct experimental basis to perform CPT tests with the BSR without assumptions about invisible decay modes.

\acknowledgments
\hspace{1.5em}
The BESIII Collaboration thanks the staff of BEPCII and the IHEP computing center for their strong support. This work is supported in part by National Key R\&D Program of China under Contracts Nos.~2020YFA0406400, 2023YFA1606000, 2020YFA0406300; the Chinese Academy of Sciences (CAS) under Contract No. U1832207; National Natural Science Foundation of China (NSFC) under Contracts Nos. 11635010, 11735014, 11935015, 11935016, 11935018, 12025502, 12035009, 12035013, 12061131003, 12192260, 12192261, 12192262, 12192263, 12192264, 12192265, 12221005, 12225509, 12235017, 12361141819;  the CAS Large-Scale Scientific Facility Program; the CAS Center for Excellence in Particle Physics (CCEPP); Joint Large-Scale Scientific Facility Funds of the NSFC; 100 Talents Program of CAS; The Institute of Nuclear and Particle Physics (INPAC) and Shanghai Key Laboratory for Particle Physics and Cosmology; German Research Foundation DFG under Contracts Nos. FOR5327, GRK 2149; Istituto Nazionale di Fisica Nucleare, Italy; Knut and Alice Wallenberg Foundation under Contracts Nos. 2021.0174, 2021.0299; Ministry of Development of Turkey under Contract No. DPT2006K-120470; National Research Foundation of Korea under Contract No. NRF-2022R1A2C1092335; National Science and Technology fund of Mongolia; National Science Research and Innovation Fund (NSRF) via the Program Management Unit for Human Resources \& Institutional Development, Research and Innovation of Thailand under Contracts Nos. B16F640076, B50G670107; Polish National Science Centre under Contract No. 2019/35/O/ST2/02907; Swedish Research Council under Contract No. 2019.04595; The Swedish Foundation for International Cooperation in Research and Higher Education under Contract No. CH2018-7756; U. S. Department of Energy under Contract No. DE-FG02-05ER41374

\newpage
\noindent M.~Ablikim$^{1}$, M.~N.~Achasov$^{4,c}$, P.~Adlarson$^{76}$, O.~Afedulidis$^{3}$, X.~C.~Ai$^{81}$, R.~Aliberti$^{35}$, A.~Amoroso$^{75A,75C}$, Q.~An$^{72,58,a}$, Y.~Bai$^{57}$, O.~Bakina$^{36}$
% ...
\\
\vspace{0.2cm}
(BESIII Collaboration)\\
\vspace{0.2cm} {\it
$^{1}$ Institute of High Energy Physics, Beijing 100049, People's Republic of China\\
% ...
$^{3}$ Bochum  Ruhr-University, D-44780 Bochum, Germany\\
$^{4}$ Budker Institute of Nuclear Physics SB RAS (BINP), Novosibirsk 630090, Russia\\
% ...
$^{35}$ Johannes Gutenberg University of Mainz, Johann-Joachim-Becher-Weg 45, D-55099 Mainz, Germany\\
$^{36}$ Joint Institute for Nuclear Research, 141980 Dubna, Moscow region, Russia\\
% ...
$^{57}$ Southeast University, Nanjing 211100, People's Republic of China\\
$^{58}$ State Key Laboratory of Particle Detection and Electronics, Beijing 100049, Hefei 230026, People's Republic of China\\
% ...
$^{72}$ University of Science and Technology of China, Hefei 230026, People's Republic of China\\
% ...
$^{75}$ University of Turin and INFN, (A)University of Turin, I-10125, Turin, Italy; (B)University of Eastern Piedmont, I-15121, Alessandria, Italy; (C)INFN, I-10125, Turin, Italy\\
$^{76}$ Uppsala University, Box 516, SE-75120 Uppsala, Sweden\\
% ...
$^{81}$ Zhengzhou University, Zhengzhou 450001, People's Republic of China\\

\vspace{0.2cm}
$^{a}$ Deceased\\
% ...
$^{c}$ Also at the Novosibirsk State University, Novosibirsk, 630090, Russia\\
% ...

}

\end{document}